\documentclass{appolb}
\usepackage{graphicx}
\usepackage{cite}

\begin{document}
\title{Towards Time Reversal Symmetry Test with o-Ps Decays using the J-PET detector 
\footnote{3\textsuperscript{rd} Jagiellonian Symposium on Fundamental and Applied Subatomic Physics, 2019}
}

\author{Juhi Raj${}^1$\thanks{juhi.raj@doctoral.uj.edu.pl}, Kamil Dulski${}^1$ and Eryk Czerwi\'nski${}^1$\\ 
{\it\normalsize ${}^1$Institute of Physics, Jagiellonian University, Krak\'ow, Poland 30-383}} 

\maketitle
\begin{abstract}
One of the features of the triplet state of positronium (ortho-Positronium) atoms is its relatively longer lifetime when compared to the singlet states of positronium (para-Positronium) atoms. 
The most probable decay of ortho-Positronium is into three annihilation photons.
In order to test the discrete symmetry using the time-reversal symmetry odd-operator, it is important to identify ortho-Positronium decay.
Identification of the decay of ortho-Positronium atoms by measuring the positronium annihilation lifetime with the Jagiellonian-Positron Emission Tomograph (J-PET) is presented in this article.

\end{abstract}
\PACS{}

\section{Introduction}
Positron Annihilation Lifetime Spectroscopy (PALS) is one of the most recognised tool for micro-structure investigation and its applications~\cite{Cassidy_1, Cassidy_2}.
This spectroscopy technique has been a foundation in identifying the long-lived radiative processes from the short-lived components in the decay of positronium atoms~\cite{Kamil}.
Recently a method combining PALS and Positron Emission Tomography was proposed and elaborated for the application in medical diagnostics\cite{P_Moskal_Nature_2019, P_Moskal_PMB_2019}.
This technique also aids the study of various physics phenomenon such as: quantum entanglement~\cite{Hiesmayr, Hiesmayr_2017}, search for mirror photons~\cite{Mirror, Bass} and test for discrete symmetries in the charged leptonic sector~\cite{Moskal_Symm}.
Time-reversal symmetry is one of the most intriguing and challenging discrete symmetries and has not been observed in the charged leptonic sector, so far~\cite{Sozzi, T2K}.
The Standard Model predicts photon-photon interaction or weak interactions to mimic the symmetry violation in the order of 10\textsuperscript{-9}~(photon-photon interaction) and 10\textsuperscript{-13}~(weak interactions), respectively~\cite{T2K, Bernreyther, Arbic, Pokraka, Bass}.
There is 6 orders of magnitude difference between the present experimental upper limit and the Standard Model predictions~\cite{Kostelecky,Yamazaki,Vetter}.
Hence, the time-reversal symmetry was proposed to be tested with the ortho-positronium (o-Ps) system by determining the expectation value of the operator constructed with the polarization $(\vec{\epsilon_i})$ and momentum direction $(\vec{k_j})$ of the annihilation photons originating from the decay of o-Ps atoms, as listed in Table 1~\cite{Moskal_Symm} using the J-PET detector.
The observation of a non-zero expectation value of this operator would imply non-invariance of these symmetries~\cite{Moskal_Symm, Yamazaki}. 

\begin{table}[h!]
\centering
\caption{Discrete symmetry odd-operator constructed using the polarization $(\vec{\epsilon_i})$ and momentum $(\vec{k_j})$ directions of the annihilation photons from the decay of o-Ps. 
The polarization direction $(\vec{\epsilon_i})$ is constructed as the cross product of the momentum direction of the primary photon $(\vec{k_j})$ and its corresponding secondary scattered photon $(\vec{k'_j})$.
The descending momentum of the three primary annihilation photons are denoted by $\mid\vec{k_i}\mid >  \mid\vec{k_j}\mid > \mid\vec{k_k}\mid$.
Symmetries for which the given operator is odd~(marked $"-"$) can be tested with the J-PET system, namely P, T and CP.
} 

\vspace{0.5cm}
\begin{tabular}{|l|l|l|l|l|l|}
\hline
\textbf{Operator}          & \textbf{C} & \textbf{P} & \textbf{T} & \textbf{CP} & \textbf{CPT} \\ 
\hline
$\vec{\epsilon_i}\cdot\vec{k_j}$        & +  & $-$  & $-$  & $-$ & +    \\
\hline
\end{tabular}
\end{table}

\section{Lifetime of ortho-Positronium atoms}
\begin{figure}[h!]
\includegraphics[width=0.48\textwidth]{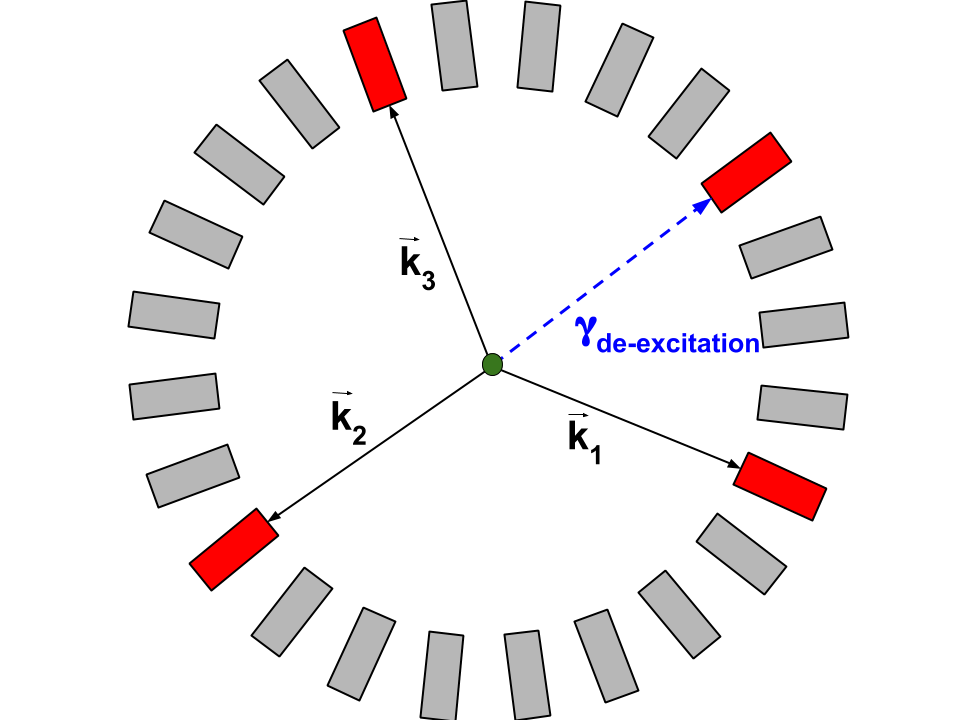}
\includegraphics[width=0.56\textwidth]{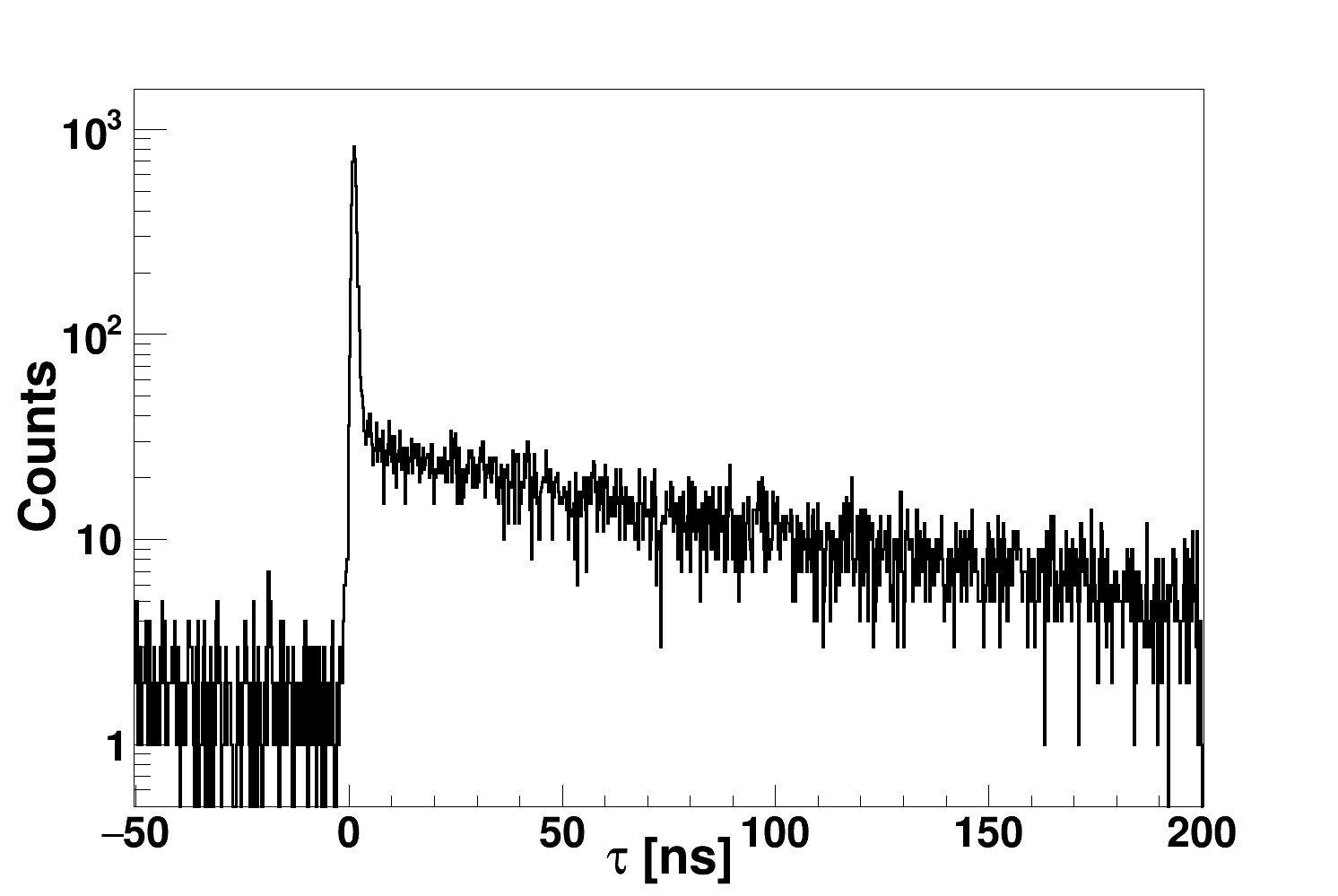}
\caption{Left Panel: Schematic of the J-PET scanner registering signal candidates, de-excitation photon($\gamma_{de-excitation}$) represented using dashed blue line and three primary annihilation photons 
$(|\vec{k_{1}}| >|\vec{k_{2}}| > |\vec{k_{3}}|)$. \newline
Right Panel: Positronium lifetime ($\tau$) distribution in the XAD-4 porous polymer, obtained from  the measurement with the J-PET detector. The lifetime distribution is obtained by identifying and time stamping the de-excitation photon and the corresponding three annihilation photons from the decay of ortho-Positronium.}
\label{fig::fig1}
\end{figure}

The J-PET detector is the first PET-scanner constructed using plastic scintillator strips to make the tomograph cost-effective and portable~\cite{Moskal_PMB_16, Moskal_NIM_14, Szymon, Greg}. 
The fast signal creation and propagation with plastic scintillators allows to time stamp the start and stop of various radiative processes. 
This helps the selection of signal events with one prompt de-excitation photon and three primary annihilation photons from the decay of ortho-Positronium as shown in the left panel of Fig.\ref{fig::fig1}

At the beginning of this year, the J-PET collaboration conducted a test experiment of producing the ortho-Positronium signal candidates by using a \textsuperscript{22}Na source placed in the center of the detector geometry covered in a porous polymer (XAD-4). 
The lifetime of the produced positronium in XAD-4 for each event is measured as the difference of emission time of the de-excitation photon and an average emission time of the three annihilation photons as shown in the right panel of Fig.~\ref{fig::fig1}~\cite{Kamil}.
The spectrum shows a sharp maximum corresponding to the annihilation of para-Positronium and direct $e+e-$ annihilations.
The long tail corresponds to the decays of ortho-Positronium atoms.
The existence of ortho-Positronium signal candidates in the data sample acts as a foundation for testing the time-reversal symmetry. 

\section{Conclusion}
The relatively long lifetime of ortho-Positronium atoms acts as a prominent feature to identify them.   
The J-PET detector is a novel PET scanner made of plastic scintillators that offer fast timing properties in order to identify the decay of ortho-Positronium atoms by estimating its lifetime. 

\section{Acknowledgement}
This work was supported by the National Science Center of Poland, through OPUS 11 with grant No.~2016/21/B/ST2/01222, the Foundation of Polish Science through the TEAM POIR.04.04.00-00-4204/17 programme and the Jagiellonian University DSC grants with numbers 2019-N17/MNS/000036 and 2019-N17/MNS/000017.

\end{document}